\documentclass[times]{elsarticle}

\usepackage[T1]{fontenc}
\usepackage{amsmath}
\usepackage{amsfonts}
\usepackage{amssymb}
\usepackage{graphicx}
\usepackage{mmacells}
\mmaSet{moredefined={LinApart, GaussianInteger, PreCollect, ApplyAfterPreCollect, GatherByDenominator}}

\usepackage[T1]{fontenc}

\def\beq{\begin{equation}}
\def\eeq{\end{equation}}
\def\bsp#1\esp{\begin{split}#1\end{split}}
\def\bal#1\eal{\begin{align}#1\end{align}}

\DeclareMathOperator*{\res}{Res}

\biboptions{sort&compress}

\begin{document}
\begin{frontmatter}

\title{{\tt LinApart}: optimizing the univariate partial fraction decomposition}

\author[1]{B.~Chargeishvili}
\ead{bakar.chargeishvili@desy.de}

\author[1,2,3]{L.~Fek\'esh\'azy\corref{cor1}}
\ead{levente.fekeshazy@desy.de}

\author[2]{G.~Somogyi} 
\ead{somogyi.gabor@wigner.hun-ren.hu}

\author[2]{S.~Van Thurenhout} 
\ead{sam.van.thurenhout@wigner.hun-ren.hu}

\cortext[cor1]{Corresponding author}

\affiliation[1]{organization={II.~Institut f\"ur Theoretische Physik, Universit\"at Hamburg}, 
	addressline={Luruper Chaussee 149}, 
	postcode={22761}, 
	city={Hamburg}, 
	country={Germany}}

\affiliation[2]{organization={HUN-REN Wigner Research Centre for Physics}, 
	addressline={Konkoly-Thege Mikl\'os \'ut 29-33},
	postcode={1121}, 
	city={Budapest}, 
	country={Hungary}}
	
\affiliation[3]{organization={Institute for Theoretical Physics, ELTE E\"otv\"os Lor\'and University}, 
	addressline={P\'azm\'any P\'eter s\'et\'any 1/A}, 
	postcode={1117}, 
	city={Budapest}, 
	country={Hungary}}	

\begin{abstract}
We present {\tt LinApart}, a routine designed for efficiently performing the univariate partial fraction decomposition of large symbolic expressions. Our method is based on an explicit closed formula for the decomposition of rational functions with fully factorized denominators. We provide implementations in both the {\sc Wolfram Mathematica} and {\sc C} languages, made available at \url{https://github.com/fekeshazy/LinApart}. The routine can provide very significant performance gains over available tools such as the {\tt Apart} command in {\sc Mathematica}.
\end{abstract}

\begin{keyword}
partial fraction decomposition
\end{keyword}

\newpageafter{abstract}

\end{frontmatter}

\section{Introduction}

Partial fraction decomposition is a standard tool commonly employed in many aspects of 
perturbative quantum field theory (QFT) calculations. Its uses include simplifying complicated 
expressions as well as bringing them to a unique form for further manipulation. In particular, it is an 
important step during the analytic computation of loop and phase space integrals. One 
possible approach to such calculations relies on deriving multidimensional real Euler-type 
integral representations for the integrals of interest and sequentially performing the integration 
over each variable in terms of functions defined as iterated integrals, such as multiple polylogarithms \cite{Goncharov:1998kja,Remiddi:1999ew,Goncharov:2001iea,Anastasiou:2013srw,Duhr:2014woa}. 
Here partial fraction decomposition of the expression in the integration variable is necessary in 
order to cast the integral into a form that can be recognized as a multiple polylogarithm. Partial fraction 
decomposition as well as its multivariate generalizations have also been applied in conjunction with the 
integration-by-parts method for obtaining simplified forms of intermediate and final results \cite{Feng:2012iq,Bendle:2019csk,Boehm:2020ijp,Badger:2021imn,Badger:2021nhg,Heller:2021qkz}.

Univariate partial fraction decomposition is of course a well-understood problem and various 
algorithms for performing such a decomposition are known, with implementations in many computer 
algebra systems\footnote{Dating back to the 1960's with Veltman's {\sc Schoonschip} \cite{Strubbe:1974vj}.}, see e.g.~\cite{Wang81,MAHONEY1983247,Book,kim:2016pf}. However these days, the complexity 
of perturbative QFT problems has reached a point where in order to obtain results, the efficiency of 
each step of the calculation must be carefully considered. In particular, the direct symbolic computation 
of phase space integrals relevant for building subtraction schemes beyond next-to-leading order can 
lead to expressions where the (lack of) efficiency of standard partial fraction decomposition routines 
becomes a bottleneck to performing the calculation \cite{Somogyi:2008fc,Bolzoni:2010bt,Somogyi:2013yk,DelDuca:2013kw}.

Hence it is desirable to have a very efficient way of performing the partial fraction decomposition. In this 
paper we present {\tt LinApart}, a univariate partial fraction decomposition routine and provide implementations in the {\sc Wolfram Mathematica} and {\sc C} languages. The routine is based on a simple closed formula following from the residue theorem and leads to massive performance gains over readily available tools such as the {\tt Apart} command in {\sc Mathematica}. This allows one to obtain solutions for a whole range of decomposition problems that were previously intractable. Moreover, the {\sc C} implementation can be linked to other computer algebra systems, such as {\sc FORM} \cite{Vermaseren:2000nd,Kuipers:2012rf}, that currently lack built-in partial fraction decomposition capabilities.

The paper is organized as follows. 
In section~\ref{sec:formula} we present our basic formula for univariate partial 
fraction decomposition. In section~\ref{sec:implementation} we discuss the implementation and usage of the 
{\tt LinApart} routine in {\sc Wolfram Mathematica} and {\sc C}. Then, in section~\ref{sec:Preformance} we 
highlight the massive performance improvements of the {\sc Mathematica} implementation with respect to {\tt Apart} 
on various classes of rational functions. Finally, in section~\ref{sec:Conclusions} we present our 
conclusions and outlook.

\section{Closed formula for univariate partial fraction decomposition}
\label{sec:formula}

Consider a rational function of the variable $x$,\footnote{Here and in the following we will denote the decomposition variable by $x$.}
\beq
R(x) = \frac{P(x)}{Q(x)}\,,
\eeq
where $P(x)$ and $Q(x)$ are polynomials of degree $\deg P$ and $\deg Q$. Obviously $R(x)$ is just 
a linear combination of monomials of $x$ divided by $Q(x)$,
\beq
R(x) = \sum_{l=0}^{\deg P} p_l \frac{x^l}{Q(x)}\,,
\eeq
hence we may restrict our attention to rational functions of the form $f(x) = \frac{x^l}{Q(x)}$. If $f(x)$ 
is a proper rational function, i.e., $l < \deg Q$, it is well-known that $f(x)$ can be written as
\beq
f(x) 
	= \sum_{i=1}^{n} \left(\frac{c_{i1}}{x-a_i} + \frac{c_{i2}}{(x-a_i)^2} + \cdots 
	+ \frac{c_{im_i}}{(x-a_i)^{m_i}}\right)\,,
\eeq
where the index $i$ counts the $n$ distinct roots $\{a_i\}_{i=1}^{n}$ of the polynomial 
$Q(x)$ and $m_i$ denotes the multiplicity of the $i$-th root. Clearly we are considering the partial 
fraction decomposition of $f(x)$ over the complex numbers, such that the polynomial $Q(x)$ can be 
written as a product of linear factors (in $x$) with positive powers: 
$Q(x) = \prod_{i=1}^{n} (x-a_i)^{m_i}$.\footnote{We note that the 
complete factorization of the denominator to linear factors is necessary in applications related to 
symbolic integration in terms of multiple polylogarithms, since multiple polylogarithms have linear 
integration kernels.} Then, the uniqueness of the Laurent series implies that $c_{ij}$ is simply the coefficient 
of the term $(x-a_i)^{-1}$ in the Laurent expansion of the auxiliary function 
$g_{ij}(x) = (x-a_i)^{j-1}f(x)$ around the point $a_i$. In other words, $c_{ij}$ is just the residue of 
$g_{ij}(x)$ at $a_i$. This residue can be directly computed as
\beq
c_{ij} = \res(g_{ij},a_i) = \frac{1}{(m_i - j)!} \lim_{x\to a_i} \frac{d^{m_i-j}}{dx^{m_i-j}}
	\Big((x-a_i)^{m_i} f(x)\Big)\,.
\label{eq:cij_Res}
\eeq
Since
\beq
(x-a_i)^{m_i} f(x) = x^l \prod_{\substack{k=1 \\ k \ne i}}^{n} \frac{1}{(x-a_k)^{m_k}}
\eeq
is independent of $a_i$, there is no subtlety in taking the limit and one may simply replace $x\to a_i$ in 
eq.~(\ref{eq:cij_Res}) to obtain an expression for $c_{ij}$ directly in terms of the roots,
\beq
c_{ij} = \frac{1}{(m_i - j)!} \frac{d^{m_i-j}}{da_{i}^{m_i-j}}
	a_{i}^l \prod_{\substack{k=1 \\ k \ne i}}^{n} \frac{1}{(a_i-a_k)^{m_k}}\,.
\eeq
Hence, $f(x)$ can be expressed as
\beq
f(x) = \sum_{i=1}^{n}\sum_{j=1}^{m_i} \frac{c_{ij}}{(x-a_i)^{j}}
	= \sum_{i=1}^{n}\sum_{j=1}^{m_i} \frac{1}{(x-a_i)^{j}} 
	\frac{1}{(m_i - j)!} \frac{d^{m_i-j}}{da_{i}^{m_i-j}}
	a_{i}^l \prod_{\substack{k=1 \\ k \ne i}}^{n} \frac{1}{(a_i-a_k)^{m_k}}\,.
\label{eq:fres1}
\eeq
This formula can be written in a more compact form by noting that the factor of $(x-a_i)^{j}$ is 
related to the $(j-1)$-st derivative of $(x-a_i)^{-1}$ with respect to $a_i$,
\beq
\frac{1}{(x-a_i)^j} = \frac{1}{(j-1)!}\frac{d^{j-1}}{d a_i^{j-1}} \frac{1}{x-a_i}\,.
\label{eq:derrule1}
\eeq
Substituting eq.~(\ref{eq:derrule1}) into eq.~(\ref{eq:fres1}), one sees that the summation over 
$j$ simply corresponds to the general Leibniz rule for the $(m_i-1)$-st derivative of a product of two 
functions,
\beq
f(x) = \sum_{i=1}^{n} \frac{1}{(m_i-1)!} \frac{d^{m_i-1}}{da_{i}^{m_i-1}}
	\left(\frac{a_{i}^l}{x-a_i} \prod_{\substack{k=1 \\ k \ne i}}^{n} \frac{1}{(a_i-a_k)^{m_k}}\right)\,.
\label{eq:fresfin}
\eeq 

Eq.~(\ref{eq:fresfin}) is in a form which can be used directly in any high-level programming 
language in which symbolic differentiation is implemented, such as 
{\sc Wolfram Mathematica}. However, it is also straightforward to perform the differentiation in 
eq.~(\ref{eq:fresfin}) symbolically, leading to an expression that involves only elementary arithmetic operations and is hence more suitable for implementation in low-level programming languages, such as {\sc C}. 
Indeed, using the multinomial generalization of the Leibniz rule,
\beq
\frac{d^m}{dx^m} \prod_{j=1}^{n} f_j(x) = 
	\sum_{j_1 + \cdots + j_n = m} \binom{m}{j_1 \ldots j_n} \prod_{l=1}^{n} \frac{d^{j_l}f_l(x)}{dx^{j_l}}\,,
\eeq
(here $\binom{m}{j_1 \ldots j_n} =\frac{m!}{j_1! \ldots j_n!}$ is a multinomial coefficient and the sum runs 
over all values of $j_1\,,\ldots \,, j_n$ that sum to $m$) we find
\beq
f(x) = \sum_{i=1}^{n} \sum_{j_{-1}+j_{0}+j_1+\cdots+\hat{j}_i+\cdots+j_n=m_i-1}
	\binom{l}{j_{-1}} \frac{a_i^{l-j_{-1}}}{(x-a_i)^{j_0+1}}
	\prod_{\substack{k=1 \\ k\ne i}}^{n} \binom{m_k+j_k-1}{j_k} \frac{(-1)^{j_k}}{(a_i-a_k)^{m_k+j_k}}\,,
\label{eq:fresfin2}
\eeq
where the hat on the index $\hat{j}_i$ serves to denote the fact that $j_i$ is missing from the set of 
indices.

Finally, we must also deal with improper rational functions of the form $f(x) = \frac{x^l}{Q(x)}$, where 
$l \ge \deg Q$. A straightforward way to proceed would be to write $f(x)$ as the sum of a polynomial 
and a proper rational function using polynomial division, after which the proper rational function part 
can be decomposed using eq.~(\ref{eq:fresfin}). However, it turns out to be more efficient to perform the 
polynomial division symbolically in the following way. First, write
\beq
f(x) = x^{l-(m-1)} \frac{x^{m-1}}{Q(x)}\,,\qquad l \ge m\,,
\label{eq:fdecomp}
\eeq
where $m=\deg Q(x) = \sum_{i=1}^{n}m_i$ is simply the degree of the denominator. Then, the second 
factor is a proper rational function by construction and we can apply the formula in eq.~(\ref{eq:fresfin}) 
to decompose it into partial fractions. After this decomposition, only rational functions of the form 
$g(x) = \frac{x^p}{(x-a)^q}$ remain. One can then perform the polynomial division symbolically as follows,
\beq
g(x) = \frac{x^p}{(x-a)^q} =
	\sum_{i=0}^{p-q} \binom{p}{i} a^i (x-a)^{p-q-i}
	+ \sum_{i=p-q+1}^{p} \binom{p}{i} a^i (x-a)^{p-q-i}\,.
\label{eq:polydiv1}
\eeq
The first term is nonzero only if $p-q\ge 0$ (otherwise this sum is empty) and 
represents the quotient polynomial, while the second term gives the proper rational function remainder 
and is in a decomposed form already. We note in passing that it is possible to write the quotient polynomial 
explicitly, since one can show that
\beq
\sum_{i=0}^{p-q} \binom{p}{i} a^i (x-a)^{p-q-i} = \sum_{i=0}^{p-q} \binom{p-1-i}{q-1}a^{p-q-i}x^i\,.
\eeq
In our actual implementation, we prefer to use this second form, i.e.,
\beq
g(x) = \sum_{i=0}^{p-q} \binom{p-1-i}{q-1}a^{p-q-i}x^i
	+ \sum_{i=p-q+1}^{p} \binom{p}{i} a^i (x-a)^{p-q-i}\,.
\label{eq:polydivfin}
\eeq

Eqs.~(\ref{eq:fresfin2})~and~(\ref{eq:polydivfin}) thus give explicit formulae representing the partial 
fraction decomposition of univariate rational functions. As noted above, for the purposes of actual 
implementation, eq.~(\ref{eq:fresfin}) is also useful in high-level languages where symbolic differentiation 
is available.

\section{Implementation and usage}
\label{sec:implementation}

In this section, we present our implementations of the {\tt LinApart} routine in the {\sc Wolfram Mathematica} and {\sc C} languages, which can be obtained at \url{https://github.com/fekeshazy/LinApart}. 

\subsection{The {\sc Wolfram Mathematica} routine}
\label{ssec:Mathcode}

The {\sc Mathematica} implementation is contained in a single file {\tt LinApart.m}. The routine can handle both proper and improper fractions. For proper fractions, the decomposition is directly computed according to eq.~(\ref{eq:fresfin}). Extensive testing has revealed that this symbolic approach, leveraging the efficiency of the built-in differentiation routine, outperforms alternative methods. For improper fractions eq.~(\ref{eq:fdecomp}) is first employed to decompose each term into a product of a monomial and a proper rational function, followed by the application of eq.~(\ref{eq:polydivfin}). By default, the resulting decomposition is returned without any further simplification or gathering of terms. However, if present, coefficients which are independent of the decomposition variable are distributed over the decomposed form using the built-in {\tt Distribute} command.

The derivations of the basic formulae we use assume that the denominator is in a fully factorized form, i.e., each denominator factor is linear in $x$, and that each multiplicity $m_i$ is a positive integer. Nevertheless, the implementation accepts any valid {\sc Mathematica} expression as input, including sums of several functions, non-linear factors in the denominator, non-integer exponents and functions of $x$ that are not rational. Sums in the input are treated simply by applying partial fraction decomposition term-by-term. Next, each term is split into a true rational function part with linear denominators and a left-over piece, which contains all non-linear denominators, factors with non-integer exponents and non-rational functions of $x$. For the purposes of this splitting, linear denominators with rational number exponents are treated as follows. First we write
\beq
(x-a)^{p/q} = (x-a)^{\lfloor p/q \rfloor} 
\cdot (x-a)^{p/q-\lfloor p/q \rfloor} \,,
\label{eq:floorpq}
\eeq
where $\lfloor r\rfloor$ denotes the floor function, i.e., the greatest integer less than or equal to $r$. The exponent of the first factor is an integer by construction, and so this factor is included in the rational function part. On the other hand, the exponent of the second factor is a rational number, strictly smaller than one and this factor enters the left-over piece. This prescription is adopted in order to reproduce the behaviour of the {\tt Apart} command on such expressions. The true rational function part is then decomposed as explained above, while the left-over piece is treated as if it were an $x$-independent overall factor.

In some practical applications, the partial fraction decomposition must be performed on large expressions with many terms. In such cases, significant performance gains can be obtained by processing the input prior to the decomposition. Examples of such operations include
    \begin{itemize}
        \item Factorization: prior to decomposition, individual terms of the input can be factored over either the integers or Gaussian integers, depending on the desired domain. Thus, non-linear denominators that factorize over these domains will also be considered during the decomposition process.        
        \item Collection of terms before decomposition: terms with identical $x$-dependence can be grouped together, reducing the number of individual decompositions required.
    \end{itemize}
These pre-processing operations are made available in our implementation as options, to be introduced below. By incorporating these features, the routine provides a robust and versatile framework for partial fraction decomposition, allowing the user to optimize performance and tailor the process to their specific needs.

The {\sc Mathematica} routine can be loaded with
\begin{mmaCell}{Code}
Needs["LinApart`"]
\end{mmaCell}
from any directory, provided the {\tt LinApart.m} file is placed in the standard {\sc Mathematica} packages directory. Alternatively, one may specify the complete path to the file when loading,
\begin{mmaCell}{Code}
Import["/Complete_Path_To_File/LinApart.m"]
\end{mmaCell}

The main function provided is {\tt LinApart}. The command {\tt LinApart[}{\it expr, var}{\tt ]} returns the partial fraction decomposition of {\it expr} with respect to the variable {\it var},
\begin{mmaCell}{Code}
\mmaDef{LinApart}[1/((1 + x)(2 + x)(3 + x)), x]
\end{mmaCell}
\begin{mmaCell}{Output}
\mmaFrac{1}{2(1+x)} - \mmaFrac{1}{2+x} + \mmaFrac{1}{2(3+x)}
\end{mmaCell}
Since the algorithm uses the built-in differentiation function, the variable must have the head {\tt Symbol}; if this condition is not fulfilled, an error message is generated and the input is returned unevaluated.

As explained above, {\tt LinApart} assumes that all denominator factors are linear in $x$. Thus, non-linear denominators, denominators with purely symbolic exponents, as well as non-rational functions of $x$ are by default ignored during decomposition.\footnote{To be more precise, the factors selected for decomposition are required to be linear polynomials in the given variable, i.e., expressions of the form $ax+b$ with $a$ and $b$ independent of $x$. Thus denominators like $\frac{1}{x+\ln x}$ are also ignored.} This implies that for expressions involving non-linear denominators, the output will not be in a completely decomposed form,
\begin{mmaCell}{Code}
\mmaDef{LinApart}[x^p Log[x]/((1 + x) (2 + x) (3 + x) (1 + x^2)), x]
\end{mmaCell}
\begin{mmaCell}{Output}
\mmaFrac{\mmaSup{x}{p} Log[x]}{2(1+x)(1+\mmaSup{x}{2})} - \mmaFrac{\mmaSup{x}{p} Log[x]}{(2+x)(1+\mmaSup{x}{2})} + \mmaFrac{\mmaSup{x}{p} Log[x]}{2(3+x)(1+\mmaSup{x}{2})}
\end{mmaCell}
In such cases, one can obtain a complete decomposition by writing the non-linear denominators in a fully factorized form. In order to facilitate such manipulations, the option {\tt Factor} is provided. When set to {\tt True}, the input expression is factorized term-by-term before the partial fraction decomposition
\begin{mmaCell}{Code}
\mmaDef{LinApart}[1/((1 - a^2) (1 - x^2)), x, "Factor" -> True]
\end{mmaCell}
\begin{mmaCell}{Output}
-\mmaFrac{1}{2(1-\mmaSup{a}{2})(-1+x)} + \mmaFrac{1}{2(1-\mmaSup{a}{2})(1+x)}
\end{mmaCell}
Notice that in order to avoid any unnecessary computation, the factorization only affects the variable-dependent part. The factorization is performed using the built-in {\tt Factor} command, i.e., by default the expression is factored over the integers. Hence, factorization will not influence irreducible (over the integers) higher-order denominators, which will continue to be ignored. However, by setting the additional option {\tt GaussianIntegers} to {\tt True} in conjunction with {\tt Factor}, factorization can be extended to allow for constants that are Gaussian integers,
\begin{mmaCell}{Code}
\mmaDef{LinApart}[1/((1 + x) (1 + x^2)), x, "Factor" -> True, 
 "GaussianIntegers" -> True]   

\end{mmaCell}
\begin{mmaCell}{Output}
-\mmaFrac{\mmaFrac{1}{4}+\mmaFrac{I}{4}}{-I+x} - \mmaFrac{\mmaFrac{1}{4}-\mmaFrac{I}{4}}{I+x} + \mmaFrac{1}{2(1+x)}
\end{mmaCell}
Further application of the built-in {\tt ComplexExpand} command allows the output to be written in a manifestly real form
\begin{mmaCell}{Code}
\mmaDef{LinApart}[1/((1 + x) (1 + x^2)), x, "Factor" -> True, 
 "GaussianIntegers" -> True] // ComplexExpand   
 
\end{mmaCell}
\begin{mmaCell}{Output}
\mmaFrac{1}{2(1+x)} + \mmaFrac{1}{2(1+\mmaSup{x}{2})} - \mmaFrac{x}{2(1+\mmaSup{x}{2})}
\end{mmaCell}

Turning to expressions involving factors with rational exponents, consider the following 
example
\begin{mmaCell}{Code}
\mmaDef{LinApart}[1/((1 + x)(2 + x)^(1/2)(3 + x)^(1/3+p)), x]
\end{mmaCell}
\begin{mmaCell}{Output}
\mmaFrac{1}{2}\mmaSqrt{2+x}\mmaSup{(3+x)}{-\mmaFrac{1}{3}-p} - \mmaFrac{\mmaSup{(3+x)}{\mmaFrac{2}{3}-p}}{\mmaSqrt{2+x}} + \mmaFrac{\mmaSqrt{2+x}\mmaSup{(3+x)}{\mmaFrac{2}{3}-p}}{2(1+x)}
\end{mmaCell}
Notice that the exponents were manipulated as in eq.~(\ref{eq:floorpq}) prior to performing the partial fraction decomposition. In particular, the factors of $\frac{1}{(2 + x)^{\frac12}}$ and $\frac{1}{(3 + x)^{\frac13+p}}$ were internally rewritten as
\beq
\frac{1}{(2 + x)^{\frac12}} = \frac{1}{2 + x} \cdot \sqrt{2 + x}
\qquad\mbox{and}\qquad
\frac{1}{(3 + x)^{\frac13+p}} = \frac{1}{3 + x} \cdot (3 + x)^{\frac23-p}\,,
\notag
\eeq
with the first factors on the right hand sides entering the partial fraction decomposition. As the example demonstrates, this rewriting is not influenced by the presence of the symbolic constant $p$ in the exponent. 
In this particular example, the output is in the same form as would be produced by {\tt Apart}.\footnote{However, exceptions can occur in cases where the exponent is a more elaborate function of rational numbers and symbolic parameters. In this case, what one considers the rational part of the exponent depends on the exact form in which it is written, e.g., consider $(x+a)^{1/2+(1-p)^2}$ and $(x+a)^{3/2 - 2p + p^2}$ which differ only in that the exponent is expanded in the second form. We choose to extract the rational part without any manipulation of the exponents (e.g., $1/2$ and $3/2$ in the first and second form), which closely matches, but is not exactly identical to the prescription used by {\tt Apart}.}

Finally, we discuss some further options that are designed to optimize the handling of large expressions with many terms. One particular situation which may arise is that the same rational function appears many times in the input with different coefficients. In such cases, rather than applying the partial fraction decomposition routine on individual terms, it can be more advantageous to first gather every unique $x$-dependent structure. Then, partial fraction decomposition only needs to be performed once per structure. In order to facilitate these operations, the option {\tt PreCollect} is included. When set to {\tt True}, unique $x$-dependent structures are first gathered in the input, before any partial fraction decomposition takes place. As stated above, when generating the output, the routine distributes the $x$-independent coefficients over the decomposed expressions, so the final output will not necessarily change when the {\tt PreCollect} option is active. However, the number of internal operations, hence the timing and memory usage, can be significantly different
\begin{mmaCell}{Code}
expr = 2/((1 + x)(2 + x)) + 1/((1 - a)(1 + x)(2 + x)) 
 + a/((1 + x)(2 + x)) - a/((1 - a)(1 + x)(2 + x)) 
 - b/((1 + a)(1 + x)(2 + x)) - (a b)/((1 + a)(1 + x)(2 + x));
 
\mmaDef{LinApart}[expr, x, "PreCollect" -> True]

\end{mmaCell}
\begin{mmaCell}{Output}
\mmaFrac{2}{1+x} + \mmaFrac{1}{(1-a)(1+x)} + \mmaFrac{a}{1+x} - \mmaFrac{a}{(1-a)(1+x)} - \mmaFrac{b}{(1+a)(1+x)} 
 - \mmaFrac{a b}{(1+a)(1+x)} - \mmaFrac{2}{2+x} - \mmaFrac{1}{(1-a)(2+x)} - \mmaFrac{a}{2+x} + \mmaFrac{a}{(1-a)(2+x)} 
 + \mmaFrac{b}{(1+a)(2+x)} + \mmaFrac{a b}{(1+a)(2+x)}

\end{mmaCell}
Indeed, in the above simple case, the output is identical to what would have been obtained without the {\tt PreCollect} option, however, the timing and memory usage is already somewhat improved as may be checked e.g.\ by using the {\tt AbsoluteTiming} and {\tt MaxMemoryUsed} commands.

When the {\tt PreCollect} option is set to {\tt True}, one may further specify the option {\tt ApplyAfterPreCollect}, which takes a pure function and applies it to the $x$-independent coefficients of the unique $x$-dependent structures which were identified by {\tt PreCollect}. A typical application would be to factor such coefficients, which can be achieved by setting {\tt "ApplyAfterPreCollect" -> Factor}
\begin{mmaCell}{Code}
expr = 2/((1 + x)(2 + x)) + 1/((1 - a)(1 + x)(2 + x)) 
 + a/((1 + x)(2 + x)) - a/((1 - a)(1 + x)(2 + x)) 
 - b/((1 + a)(1 + x)(2 + x)) - (a b)/((1 + a)(1 + x)(2 + x));
 
\mmaDef{LinApart}[expr, x, "PreCollect" -> True, 
 "ApplyAfterPreCollect" -> Factor]

\end{mmaCell}
\begin{mmaCell}{Output}
\mmaFrac{3}{1+x} + \mmaFrac{a}{1+x} - \mmaFrac{b}{1+x} - \mmaFrac{3}{2+x} - \mmaFrac{a}{2+x} + \mmaFrac{b}{2+x}
\end{mmaCell}
As the above example shows, this can lead to significant simplifications. However, one must be aware that if the coefficients which arise after gathering unique structures are large, performing the factorization can become quite expensive in terms of runtime. In order to mitigate this to a certain extent, the pure function {\tt GatherByDenominator} is defined. This function gathers its input by denominators and can already lead to substantial simplifications without a complete factoring of the coefficients generated by 
{\tt PreCollect},
\begin{mmaCell}{Code}
expr = 2/((1 + x)(2 + x)) + 1/((1 - a)(1 + x)(2 + x)) 
 + a/((1 + x)(2 + x)) - a/((1 - a)(1 + x)(2 + x)) 
 - b/((1 + a)(1 + x)(2 + x)) - (a b)/((1 + a)(1 + x)(2 + x));
 
\mmaDef{LinApart}[expr, x, "PreCollect" -> True, 
 "ApplyAfterPreCollect" -> GatherByDenominator]

\end{mmaCell}
\begin{mmaCell}{Output}
\mmaFrac{3}{1+x} + \mmaFrac{a}{1+x} + \mmaFrac{-b-a b}{(1+a)(1+x)} - \mmaFrac{3}{2+x} - \mmaFrac{a}{2+x} - \mmaFrac{-b-a b}{(1+a)(2+x)}
\end{mmaCell}
In this case, the simplification is not complete, but nevertheless, the output is already in a much simpler form than without the use of the 
{\tt ApplyAfterPreCollect} option. It is important to note though, that the choice of the appropriate {\tt ApplyAfterPreCollect} setting is highly context-dependent and may require careful consideration of the specific characteristics of the algebraic expressions involved. The list of options is summarized in table~\ref{tab:options}.
\begin{table}
\begin{center}
\renewcommand{\arraystretch}{1.5}
\begin{tabular}{|p{0.32\textwidth}|p{0.58\textwidth}|}
\hline 
Option & Description \\
\hline\hline
{\tt Factor} & 
{\tt True}/{\tt False}: If set to {\tt True}, the input expression is factorized term-by-term before partial fraction decomposition. The default value is {\tt False}.
\\
\hline
{\tt GaussianInteger} &  {\tt True}/{\tt False}: If set to {\tt True} in conjunction with {\tt Factor}, factorization of the input expression is performed over the Gaussian integers. The default value is {\tt False}.
\\
\hline
{\tt PreCollect} &  {\tt True}/{\tt False}: If set to {\tt True}, unique variable-dependent structures are gathered in the input before partial fraction decomposition. The default value is {\tt False}.
\\
\hline
{\tt ApplyAfterPreCollect} &  pure function (e.g., {\tt Factor}): If the option {\tt PreCollect} is set to {\tt True}, the pure function specified by this option is applied to the coefficients of the unique variable-dependent structures identified by {\tt PreCollect}. Typical functions might be {\tt Factor} or {\tt GatherByDenominator}. By default, this option is empty (no function is applied to the coefficients).
\\
\hline
\end{tabular}
\end{center}
\caption{\label{tab:options} The list of options for the {\tt LinApart} command.}
\end{table}

\subsection{The {\sc C} routine}
\label{ssec:Ccode}

We have also implemented our routine based on eqs.~\eqref{eq:fresfin2} and~\eqref{eq:polydivfin} in the {\sc C} programming language for rational functions of the form 
\beq
x^l \prod_{k=1}^{n} \frac{1}{(x-a_k)^{m_k}}\,,
\label{eq:cinput}
\eeq
where $l$ is a non-negative integer and $m_k$, $k=1,\ldots,n$ are strictly positive integers. Moreover, the $a_k$ are assumed to be distinct. It is packaged into a standalone executable as well as a library suitable for linking with other software. In particular we provide the header file \texttt{LinApart.h} which can be used as a developer library to create other tools and programs that require partial fraction decomposition functionality.

One of the key features of our implementation is its efficiency in terms of both time and space complexity. The GNU Multiple Precision Arithmetic Library (GMP)~\cite{gmp} library is used for all multiple-precision integer operations, such as calculations of the factorials of large numbers, while dynamic memory allocation is carefully managed. To limit memory usage, the program employs a buffering strategy where intermediate results are accumulated in a fixed-size buffer. When the buffer reaches its capacity, the contents are flushed to disk. This approach allows the program to handle very large inputs that would otherwise exceed available memory. Only the memory required by GMP to perform the arithmetic operations is consumed such that the memory usage will not exceed a couple kilobytes in all but the most extreme applications.

The standalone program can be compiled on Unix-like systems simply by running \texttt{make}, as the only external dependency is the GMP library which is widely available. After compiling the standalone version of {\tt LinApart}, one can access the program's usage instructions with the following command:
\begin{verbatim}
    ./LinApart -h
\end{verbatim}
The program expects two strings of parameters in the following order:
\begin{verbatim}
    ./LinApart <exponents> <roots>
\end{verbatim}
where {\tt <exponents>} should contain the comma separated list of the $(m+1)$ integers $l, m_1,\ldots m_n$ defined in eq.~(\ref{eq:cinput}). I.e., the first entry represents the exponent of the monomial in the numerator and the rest are the multiplicities of the factors in the denominator. {\tt <roots>} should contain the comma separated list of roots $a_1,\ldots,a_n$. The decomposition variable is automatically set to $x$. For instance, the command:
\begin{verbatim}
    ./LinApart "3,5,7,11" "a1,a2,a3"
\end{verbatim}
will perform the partial fraction decomposition of
\begin{equation*}
    \frac{x^3}{(x-a_1)^5(x-a_2)^7(x-a_3)^{11}}\, .
\end{equation*}
The result will be written to the standard output and saved in a file named {\tt result.out}. 

\section{Performance}
\label{sec:Preformance}

As stated in the Introduction, the motivation for developing {\tt LinApart} was to obtain a highly 
efficient univariate partial fraction decomposition algorithm. In this section, we examine the performance 
of our implementation with respect to the standard {\tt Apart} command of {\sc Mathematica} as 
a function of the complexity of the input. First, let us consider what makes a rational fraction ``complex''. 
Several factors come to mind immediately:
\begin{enumerate}
\item The number of distinct denominator factors.
\item The complexity of each individual denominator. In fact, even considering only linear denominators 
of the form $x-a_i$, the roots $a_i$ may be functions of further variables and symbolic constants. 
\item The multiplicity of the denominator factors.
\item The polynomial order of the numerator.
\end{enumerate}

\begin{figure}
\begin{center}
\includegraphics{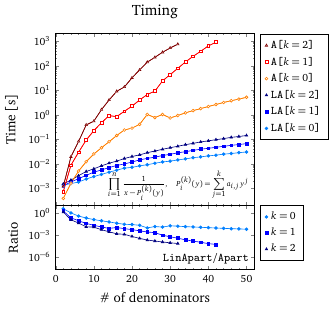}
\hspace{0.5em}
\includegraphics{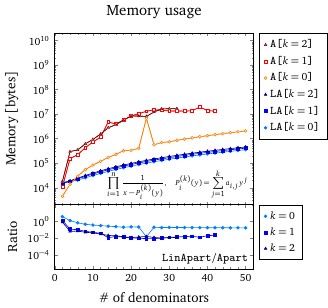}
\caption{\label{fig:RES1m1ki} Timings and memory usage of {\tt Apart} and {\tt LinApart} 
(denoted as \texttt{A} and \texttt{LA} in the legend) on rational functions of $x$ with $n$ distinct 
denominators of multiplicity one of the form $x-P^{(k)}(y)$. The roots $P^{(k)}(y)$ are chosen to be 
symbolic polynomials in the auxiliary variable $y$ of order $k$. Various curves correspond to different 
polynomial orders $k=0,1,2$ in the roots. The numerator has been set to 1.}
\end{center}
\end{figure}
These various aspects of complexity can be captured by choosing suitable input functions for the partial 
fraction decomposition routines. For example, consider the expression
\beq
\prod_{i=1}^{n} \frac{1}{x-P^{(k)}_i(y)}
\qquad{\mbox{with}}\qquad
P^{(k)}_i(y) = \sum_{j=1}^{k}a_{i,j}y^j\,.
\label{eq:testfcn1}
\eeq
Choosing the $a_{i,j}$ to be symbolic constants, we can vary the complexity by varying the total number 
of factors $n$, as well as the polynomial order $k$ of the roots. In figure~\ref{fig:RES1m1ki}, we present 
the timings and memory usage\footnote{Timing and memory usage information were obtained with the {\tt AbsoluteTiming} and {\tt MaxMemoryUsed} commands.} of {\tt Apart} and {\tt LinApart} on expressions of the form 
given in eq.~(\ref{eq:testfcn1}). During each evaluation, we have constrained the runtime in 
{\sc Mathematica} to a maximum of $10^3$ seconds. Hence, missing data points represent computations 
that did not finish with this time constraint.\footnote{We have performed these evaluations on a standard 
desktop PC with an Intel Core i5 processor and 16Gb of RAM.} The number of denominators was chosen 
between 2 and 50 and we varied the polynomial order of the roots between 0 and 2. The figure clearly 
shows that for basically all of the considered cases, {\tt LinApart} outperforms {\tt Apart} both 
in terms of timing and memory usage. Indeed, already for just six denominators, {\tt LinApart} 
produces a speedup of a factor between $\sim$2 and $\sim$20 depending on the polynomial order of 
the roots. For ten denominators, the speedup is between a factor of  $\sim$5 and $\sim$100, while for 
tens of denominators and roots of polynomial order two, we observe speedups of factors greater than 
$10^4$. On these expressions, {\tt LinApart} also typically outperforms {\tt Apart} in terms 
of memory used by factors of $\sim$10--100, depending on the number of denominators and the complexity 
of the roots.

\begin{figure}
\begin{center}
\includegraphics{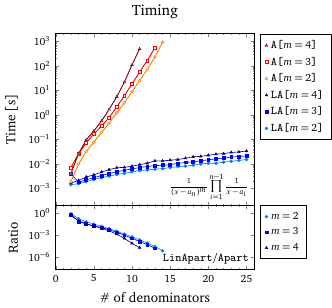}
\hspace{0.5em}
\includegraphics{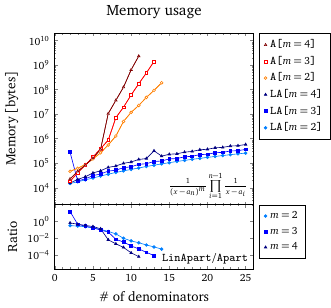}
\caption{\label{fig:RES1mik0} Timings and memory usage of {\tt Apart} and {\tt LinApart} 
(denoted as \texttt{A} and \texttt{LA} in the legend) on rational functions of $x$ with $n$ distinct 
denominators of multiplicity one, except for the last denominator, whose multiplicity is $m$. The roots 
$a_i$ are chosen to be symbolic constants. Various curves correspond to different multiplicities $m=2,3,4$ 
of the last denominator. The numerator has been set to 1.}
\end{center}
\end{figure}
Next, let us study the effect of higher multiplicities. To begin, we examine the case when 
only one out of the $n$ distinct denominators has multiplicity $m>1$,
\beq
\frac{1}{(x-a_n)^m} \prod_{i=1}^{n-1}\frac{1}{x-a_i}\,,\qquad 2\le m\,.
\label{eq:testfcn2}
\eeq
The results for timings and memory usage of {\tt Apart} and {\tt LinApart} on 
inputs of the form of eq.~(\ref{eq:testfcn2}) are presented in figure~\ref{fig:RES1mik0}. As before, 
a time constraint of $10^3$ seconds was imposed on every evaluation. The number of denominators 
was varied between 2 and 25, and the multiplicity of the last denominator between 2 and 4. Again, 
we observe a dramatic improvement in performance using {\tt LinApart}, both in terms of 
necessary time and memory. In fact, already for expressions with only three denominators, we 
obtain a speedup of a factor of $\sim$ 10. On expressions with ten denominators, the speedup is already 
greater than a factor of $10^3$. Increasing the number of denominators further, we quickly exhaust the 
cases where the evaluation with {\tt Apart} finishes in the chosen time constraint. Indeed, for 
$m=2$ this happens already at $n=14$ (i.e., just 13 denominators of multiplicity one and a single 
denominator with multiplicity two). In this case, we observe a speedup of a factor greater than $10^5$. 
Examining the memory usage, we also see significant gains of up to factors of $10^4$ in favour of 
{\tt LinApart}.

\begin{figure}
\begin{center}
\includegraphics{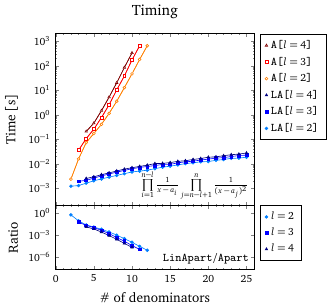}
\hspace{0.5em}
\includegraphics{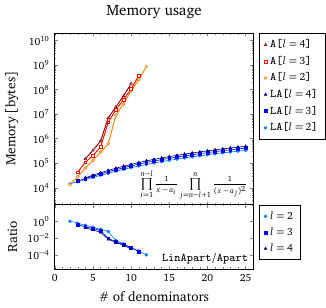}
\caption{\label{fig:RES2mik0} Timings and memory usage of {\tt Apart} and {\tt LinApart} 
(denoted as \texttt{A} and \texttt{LA} in the legend) on rational functions of $x$ with $n$ distinct 
denominators. The multiplicities of the first $n-l$ denominators are one, while the last $l$ denominators have 
multiplicity two. The roots $a_i$ are chosen to be symbolic constants. Various curves correspond to the 
different number $l=2,3,4$ of quadratic denominators . The numerator has been set to 1.}
\end{center}
\end{figure}
It is also interesting to consider the case of multiple denominators whose multiplicity is greater than 
one. Thus, in figure~\ref{fig:RES2mik0}, we examine the timings and memory usage of {\tt Apart} 
and {\tt LinApart} on expressions of the form
\beq
\prod_{i=1}^{n-l}\frac{1}{x-a_i} \prod_{j=n-l+1}^{n} \frac{1}{(x-a_j)^2}\,,\qquad 1\le l \le n\,,
\eeq
i.e., $l$ out of the $n$ distinct denominators have multiplicity two, while the rest of the $(n-l)$ denominators 
have multiplicity one. In terms of both evaluation times and memory used, we observe the same 
dramatic performance gains using {\tt LinApart} as in the previous case of a single denominator 
of higher multiplicity. Again, speedups of factors of $\sim$$10^4$--$10^5$ are obtained for expressions 
with a total number of only around ten denominators. At the same time, the required memory also decreases 
by up to four orders of magnitude.

\begin{figure}
\begin{center}
\includegraphics{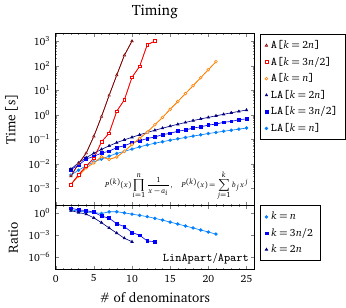}
\hspace{0.5em}
\includegraphics{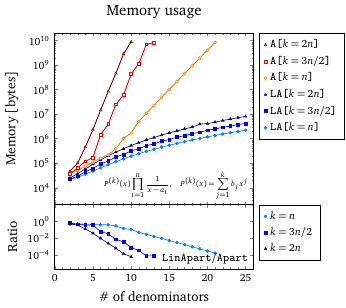}
\caption{\label{RES3m1k0li} Timings and memory usage of {\tt Apart} and {\tt LinApart} 
(denoted as \texttt{A} and \texttt{LA} in the legend) on improper rational functions of $x$ with $n$ 
distinct denominators of multiplicity one. The roots $a_i$ are chosen to be symbolic constants, while 
the numerator is a symbolic polynomial $P^{(k)}$ of $x$. Various curves correspond to  
different polynomial orders $k=n,\lfloor3n/2\rfloor,2n$.}
\end{center}
\end{figure}
Next, we consider the effects of including non-trivial numerators,
\beq
P^{(k)}(x) \prod_{i=1}^{n}\frac{1}{x-a_i}
\qquad\mbox{with}\qquad
P^{(k)}(x) = \sum_{j=1}^{k}b_j x^j\,.
\label{eq:testfunc3}
\eeq
The timings and memory usage obtained with {\tt Apart} and {\tt LinApart} on improper 
rational function inputs of the form of eq.~(\ref{eq:testfunc3}) are presented in figure~\ref{RES3m1k0li} 
for numerators with polynomial orders $k=n$, $\lfloor3n/2\rfloor$ and $2n$. Once again, huge improvements 
in performance of  {\tt LinApart} over {\tt Apart} are apparent in both timing and memory 
used. Concerning memory usage in particular, we note that during these evaluations, we had to enforce 
a memory constraint of 12Gb in order to avoid the automatic closing of the {\sc Mathematica} kernel while {\tt Apart} was running.

\begin{figure}
\begin{center}
\includegraphics{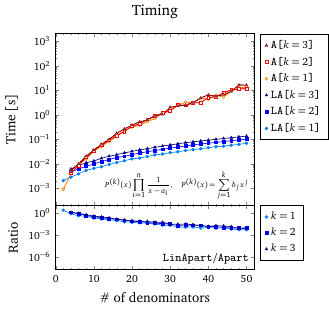}
\hspace{0.5em}
\includegraphics{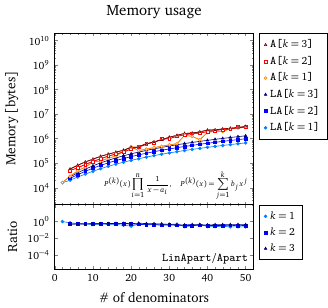}
\caption{\label{RES3m1k0li2} Timings and memory usage of {\tt Apart} and {\tt LinApart} 
(denoted as \texttt{A} and \texttt{LA} in the legend) on proper rational functions of $x$ with $n$ 
distinct denominators of multiplicity one. The roots $a_i$ are chosen to be symbolic constants, while 
the numerator is a symbolic polynomial $P^{(k)}$ of $x$. Various curves correspond to  
different polynomial orders $k=1,2,3$.}
\end{center}
\end{figure}
For proper rational functions, we present the timings and memory usage of {\tt Apart} and 
{\tt LinApart} in figure~\ref{RES3m1k0li2}. Here the total number of denominators was varied 
between 2 and 50 and we have chosen numerators with polynomial orders $k=1$, $2$ 
and $3$. As in all previous cases {\tt LinApart} outperforms {\tt Apart} except for the 
simplest cases, especially in terms of timing. Interestingly, in this case only, we do not observe very 
large (i.e., orders of magnitude) gains in terms of memory usage in favour of {\tt LinApart}.

We finish this section by presenting an example which emerges during the analytic
computation of phase space integrals relevant for setting up a local subtraction scheme beyond next-to-leading order. After choosing a particular phase space parametrization we encounter expressions of the form
\beq
\bsp
f(x_a,x_b;y) &=
\Big[(-4 + y) (1 - y + x_b y) (2 - y + x_b y) (4 - y + x_b y) (1 - x_a - y + x_b y)^3 
\\&\times
(-1 + x_a - y + x_b y) (-4 - 4 x_b - y + x_b y) (-4 x_b - y + x_b y) \\&\times
(-4 x_a - 4 x_b - y + x_b y) (4 x_a - 4 x_b - y + x_b y) (2 + 2 x_b - y + x_b y)^3 
\\&\times
(6 + 2 x_b - y + x_b y) (2 - 4 x_a + 2 x_b - y + x_b y) (2 + 4 x_a + 2 x_b - y + x_b y) 
\\&\times
(-1 + x_a - x_a y + x_a x_b y) (1 + x_a - x_a y + x_a x_b y) (-2 + 2 x_a - x_a y + x_a x_b y) 
\\&\times
(2 + 2 x_a - x_a y + x_a x_b y) (-x_b + x_a x_b - x_a y + x_a x_b y)^3 
\\&\times
(-4 + 2 x_a + 2 x_a x_b - x_a y + x_a x_b y) (4 + 2 x_a + 2 x_a x_b - x_a y + x_a x_b y) 
\\&\times
(1 - 2 x_a + x_a^2 - y - x_a y + x_b y + x_a x_b y) 
\\&\times
\left(2 x_b - 2 x_a x_b + x_a y - x_b y - x_a x_b y + x_b^2 y\right)^3\Big]^{-1}\,,
\esp
\label{eq:realex}
\eeq
which must be integrated symbolically over $y$. To do so, one must first perform the partial fraction decomposition with respect to $y$. The rational function in eq.~(\ref{eq:realex}) has 23 linear (in $y$) denominators, 4 of which have multiplicity 3. A quick glance at figure~\ref{fig:RES2mik0} shows that {\tt Apart} already requires close to $10^3$ seconds to decompose a rational function with just 10 denominators, 4 of which have multiplicity 2. A naive extrapolation to 23 denominators then puts the time required for the decomposition of $f(x_a,x_b;y)$ at the order of $10^9$ seconds.\footnote{As we are dealing now with denominators of multiplicity 1 and 3 instead of 1 and 2 as in figure~\ref{fig:RES2mik0}, we expect this number to be a rough lower limit. The situation is made even worse by the fact that eq.~(\ref{eq:realex}) contains roots which are non-trivial polynomials of $x_a$ and $x_b$. This also has a significant influence on the efficiency of {\tt Apart}. Hence we expect the actual time required to be several orders of magnitude larger than this naive estimate.}. On the other hand, {\tt LinApart} performs the partial fraction decomposition in a mere $\sim 10^{-2}$ seconds (without using any of the provided options). As expressions actually encountered in real computations typically have hundreds or even thousands of rational functions of the type in eq.~(\ref{eq:realex}), this is a vital improvement.

\section{Conclusions}
\label{sec:Conclusions}

In this paper, we have introduced the {\tt LinApart} routine, which provides a fast and efficient implementation of univariate partial fraction decomposition for rational functions with fully factorized denominators. Our implementation is based on a simple closed formula for the decomposed function and we provide realizations in the {\sc Wolfram Mathematica} and {\sc C} languages. The {\sc Mathematica} routine leverages the highly-efficient built-in differentiation routine, while our {\sc C} code can be linked to computer algebra systems such as {\sc FORM}, where symbolic differentiation is not available.

Concentrating on the {\sc Mathematica} implementation, we have found that {\tt LinApart} outperforms the built-in {\tt Apart} function on virtually any input expression both in terms of timing and memory used. The increase of efficiency is dramatic already for reasonably small expressions, especially if the multiplicities of some of the denominator factors are greater than one. Indeed, for rational functions involving only around ten independent denominators, with only a few (say 2 to 4) denominator factors of multiplicity two, speedups of up to five orders of magnitude are observed. At the same time, the required memory decreases by up to four orders of magnitude. Thus, {\tt LinApart} is able to efficiently handle decomposition problems that are intractable for the built-in routines.

A limitation of the current implementation concerns the treatment of non-linear (in the decomposition variable) denominator factors. While any polynomial can be factored into linear factors over the complex numbers, the partial fraction decomposition problem over the reals is understood to allow the appearance of quadratic denominators as well. Including such factors in our algorithm is in principle possible, but finding the most efficient way to do so is non-trivial and is left for future work.

\section*{Acknowledgements}
The authors would like to thank Sven-Olaf Moch for useful discussions. This work has been supported by grant K143451 of the National Research, Development and Innovation Fund in Hungary. Furthermore, the work of G.S. was supported by the Bolyai Fellowship program of the Hungarian Academy of Sciences. The work of L.F. was supported by the German Academic Exchange Service (DAAD) through its Bi-Nationally Supervised Scholarship program.

\newpage
\bibliographystyle{elsarticle-num}
\bibliography{LinApart}

\end{document}